\documentclass[12pt]{iopart}

\usepackage{iopams}  
\usepackage{graphicx}

\begin{document}

\title[]{Equilibrium potential profile across magnetic field in inhomogeneous electronegative plasma }

\author{Pawandeep Singh$^{1,2}$, Swati$^{1,2}$ \& Shantanu Kumar Karkari$^{1,2}$}

\address{$^{1}$Institute for Plasma Research, Bhat, Gandhinagar, Gujarat, 382428 India}
\address{$^{2}$Homi Bhabha National Institute, Training School Complex, Anushakti Nagar, Mumbai-400094, India}

\ead{singh.pawandeep67@gmail.com, dahiyaaswati@gmail.com  and  skarkari@ipr.res.in }
\vspace{10pt}

\begin{abstract}
This paper presents the plasma potential distribution across magnetic field lines in an argon/oxygen plasma created by 13.56 MHz CCRF discharge consisting of a pair of ring electrodes. It is found that the radial plasma density, electron temperature and plasma potential shows unusual trend as the magnetic field strength and gas pressure increases. An analytical model has been formulated to explain the radial plasma potential profile when the short-circuiting effect due to conducting boundaries can be neglected. It is found that with certain approximations, the plasma potential distribution can be very well related to the local plasma density, electron temperature and bulk electronegativity similar to a Boltzmann relation. 
\end{abstract}

\maketitle
\section{Introduction}

It is well known that plasmas created in laboratory devices generally tend to develop a positive potential relative to the confining boundaries due to highly mobile electrons, which tends to escape rapidly towards the grounded surface. In order to preserve charge neutrality, the plasma becomes slightly polarized relative to the external electrodes in contact with it; thereby it accelerates the positive ions and retards the electrons towards the boundary walls \cite{Bohm1949a,Mott-Smith1926a}. The knowledge about the spatial potential distribution inside a plasma system is hugely important, since it governs the overall dynamics of the charged particles from one region to another \cite{Smirnov2013,Stern1975}. As a matter of fact, the role of plasma potential is well known in the case of electrostatic probes \cite{Cherrington1982}, satellite charging in space \cite{Singh1996}, plasma processing applications, in plasma based ion sources \cite{Taccogna2010,Sonato2009} for generating particle beams and plasma wall interactions \cite{Chabert2011,Lieberman2005} in fusion devices \cite{Jacquot2014}. 

In low pressure collision-less system, the electron temperature is generally found to be isotropic in nature. As a result, the plasma density and potential distribution inside the plasma follow a Boltzmann distribution. However, it has been seen that the spatial potential distribution inside the plasma can be highly system specific as it depends on various other experimental conditions, like external magnetic field \cite{Das2019a}, the discharge source \cite{Schulze2008}, and the type of gases used \cite{Kim2014}. Numerous examples can be found in the literature, which shows significant deviation from the theoretical prediction of plasma potential with regard to the experiments \cite{Das2019a,Schulze2008,Bogdanov2015,Ida2017,Schwirzke1964}. Particularly, the magnetized plasma systems is one such example where the presence of external electrodes in contact with the plasma body are found to dramatically alter the equilibrium properties of plasma from the Boltzmann to a non-Boltzmann potential distribution profile, particularly across magnetic fields \cite{Das2019a,Schulze2008,Schwirzke1964}. In another case of electro-negative discharges, the negative ions are known to affect the plasma potential distribution between the discharge plates. It is found that the potential inside the bulk plasma is greatly reduced due to the presence of negative ions inside the bulk plasma \cite{Monahan2008,Lichtenberg2000}; whereas an opposing effect can be observed when a negative ion emitting electrode made of low work function surface \cite{Bacal2006,Pandey2020} is introduced in the plasma. In certain cases like in an ion beam source, there observes a steep gradient in electron temperature, density and plasma potential distribution. Therefore the influence of potential distribution on the charge particle transport in the magnetic filter field lines \cite{Shah2021,Kumar2004} is quite strong. It would be important to understand the nature of plasma potential distribution in such experimental plasma systems; however, it is difficult to predict whether the density and electron temperature distribution affects the plasma potential distribution or it is otherwise.

In this paper, the equilibrium plasma potential distribution across magnetic field due to the presence of spatially varying negative ion density and electron temperature in a capacitive coupled oxygen discharge; is investigated. In this experiment it was observed that the relative variation in plasma potential with regard to the central region, neither followed the plasma density nor the electron temperature profile. But it can be expressed by a modified Boltzmann formula which relates the plasma potential with the characteristics density, electron temperature as well as the bulk electronegativity parameter $\alpha$. 

Section-\ref{ExpSet}, briefly describes about the experimental setup. The experimental results are presented in section-\ref{ExpRes}, followed by the analytical model of plasma potential in section-\ref{Model}. The comparison between experiments and the modelling results are discussed in section-\ref{ResDis}. Finally the paper concludes by summarizing the main findings in section-\ref{Sum}.

\section{\label{ExpSet}Experimental Setup:}
\begin{figure*}[!hbp]
  \centering
  \begin{minipage}[h]{0.49\textwidth}
    \includegraphics[width=\textwidth]{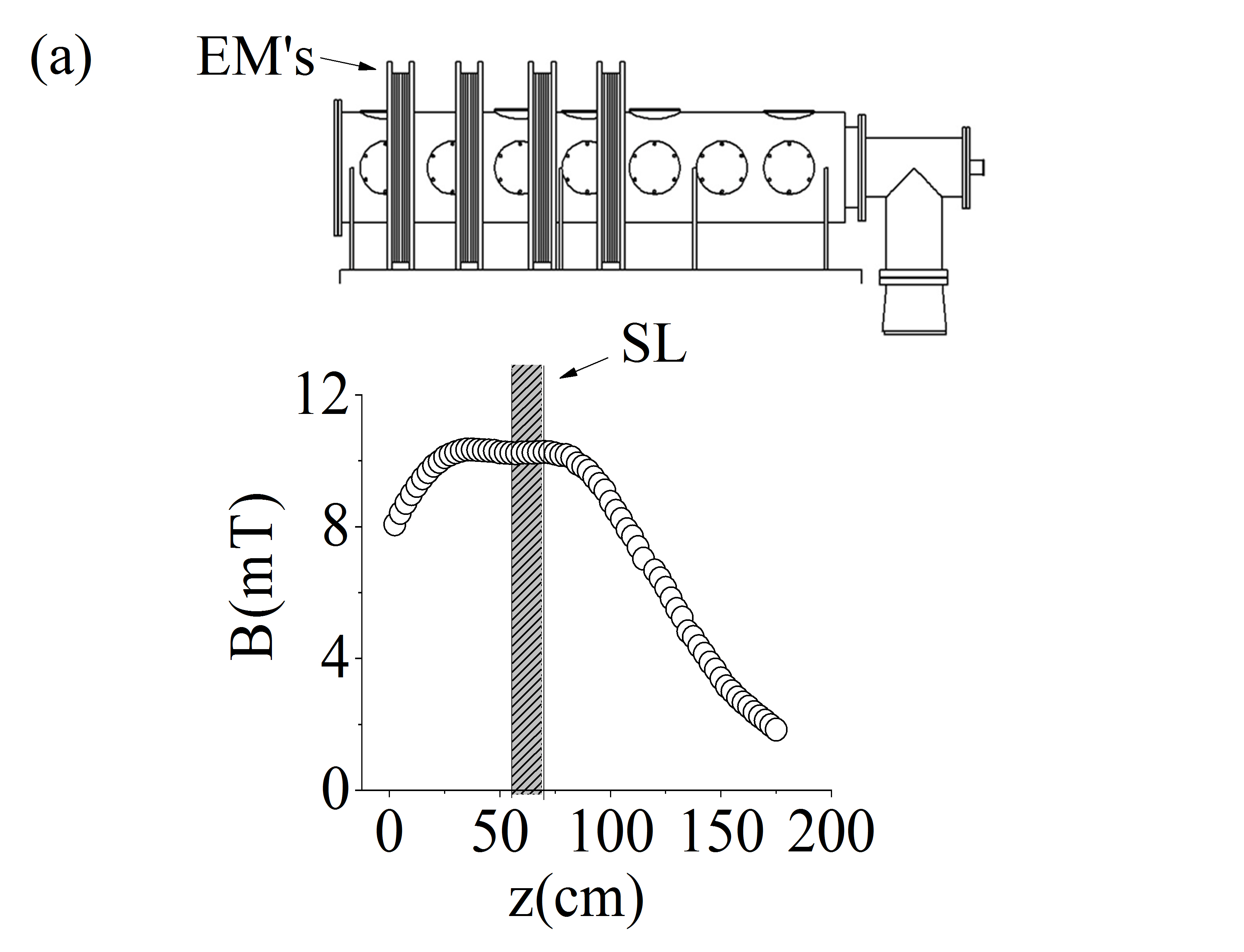}
  \end{minipage}
  \hfill
  \begin{minipage}[h]{0.49\textwidth}
    \includegraphics[width=\textwidth]{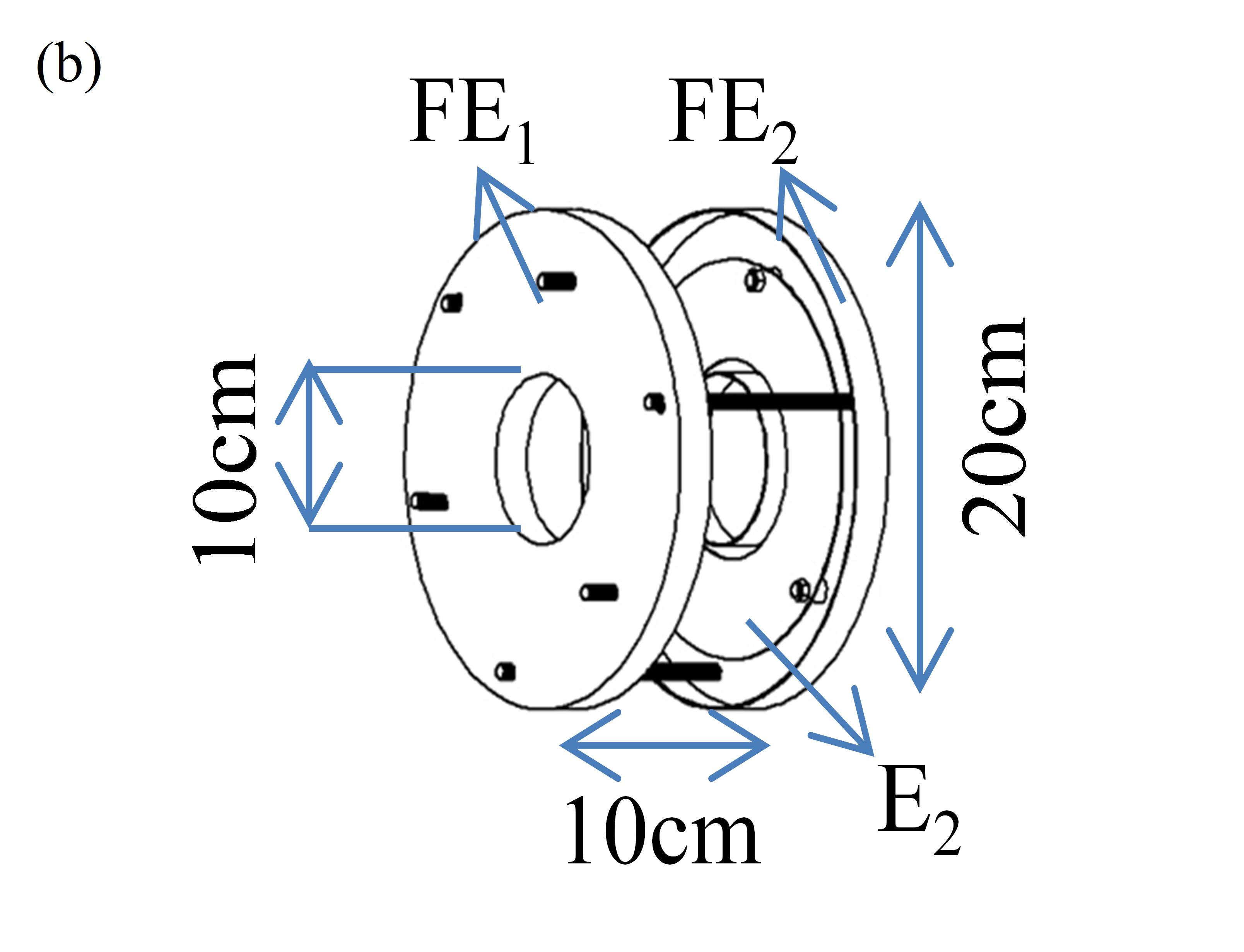}
    
  \end{minipage}
\caption{\label{ExpSet01} (a) Experimental Setup: vacuum chamber with four electromagnets (EM's) and its magnetic field configuration, SL- source location, (b) Source configuration: $E_1$ (which is in front of $E_2$) $\&$ $E_2$ are the ring electrodes, the ring electrodes are encapsulated by floating electrodes ($FE_1$ $\&$ $FE_2$), OD, ID and the gap between the $E_1$ $\&$ $E_2$ are kept to be 10cm, 20cm and 10cm respectively.}
 
\end{figure*}

\begin{figure*}[!hbp]
  \centering
     \includegraphics[width=0.65\textwidth]{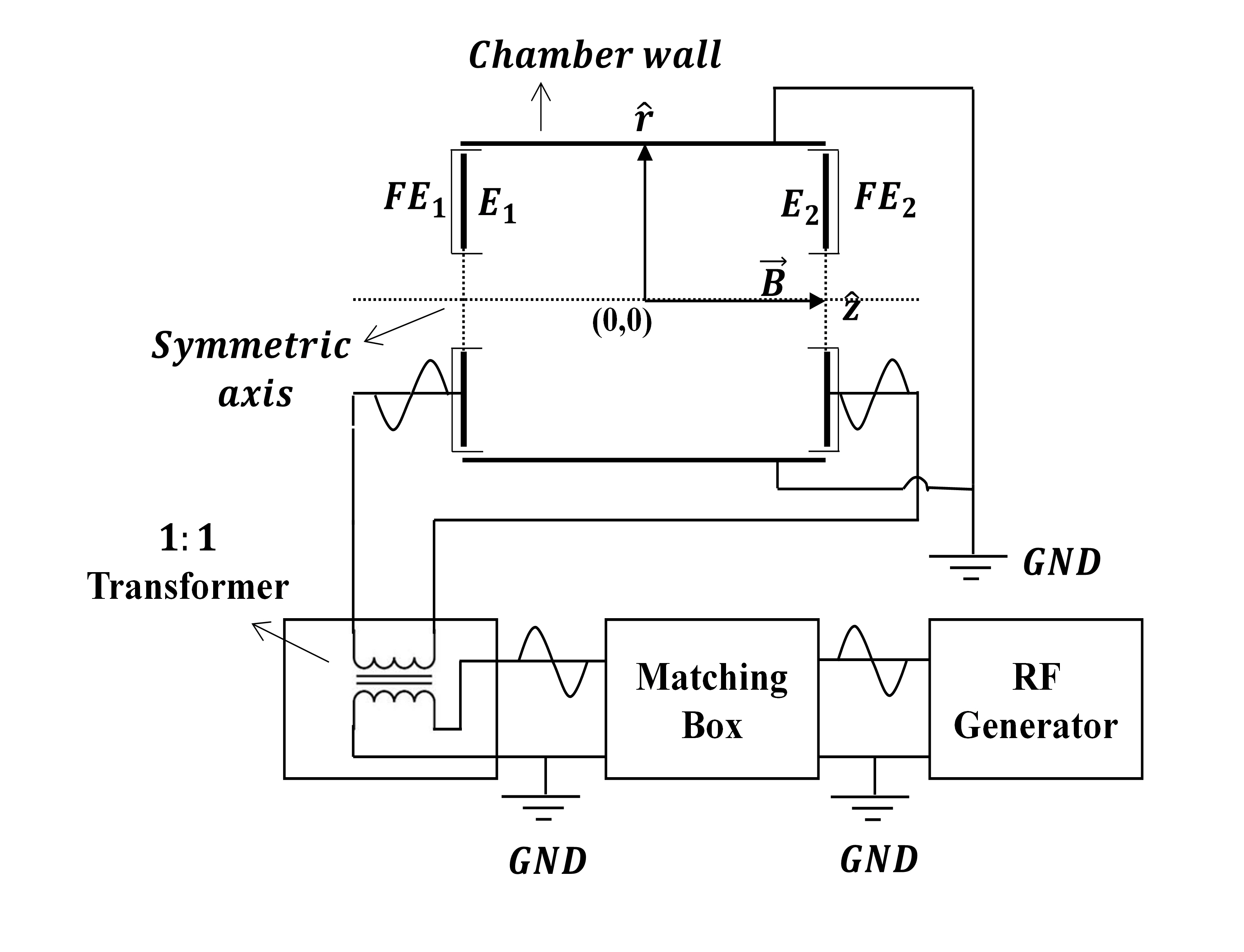}
    \caption{\label{ExpSet02} Schematic of experimental setup: $FE_1$ $\&$ $FE_2$-floating electrodes, $E_1$ $\&$ $E_2$-ring electrodes, GND-ground, symmetric axis shows the azimuthal symmetry of the source as well as vacuum chamber.}
  
\end{figure*}

In figure~\ref{ExpSet01} the experimental setup is presented. The plasma is created between a pair of axially spaced ring electrodes, which has a 5 cm inner diameter and 10 cm outer diameter. The plates are separated by a gap of 10 cm and the outer surfaces of the electrodes are encapsulated inside floating electrodes to allow the discharge to be created between the plates. All the electrodes inside the discharge setup are made of non-magnetic SS304. Four electromagnets arranged in Helmholtz configuration were used to generate a uniform magnetic field over a 28 cm diameter of the vacuum chamber. To create the discharge, a 13.56 MHz RF source with an automated matching unit [Co-Axial power system] has been used. The output from the matching unit is connected to the input of a 1:1 isolation transformer, and the output of the transformer is hooked to the floating discharge plates. This configuration provides an electrode free central region, which naturally produces a gradient in plasma density and electron temperature due to diffusion of charge particles across the axial magnetic field lines. The schematic of this configuration is also shown in figure~\ref{ExpSet02}. In the experiment both argon as well as oxygen gas has been used. The radial plasma parameters are obtained using an RF-compensated Langmuir probe, introduced at the mid-plane between the discharge plates. A precision gas dosing valve has been used to regulate the pressure inside the discharge chamber which was maintained at 2.5 mTorr.

\section{\label{ExpRes}Experimental Results:}
\begin{figure}[!hbp]
  \centering
  \begin{minipage}[b]{0.49\textwidth}
    \includegraphics[width=\textwidth]{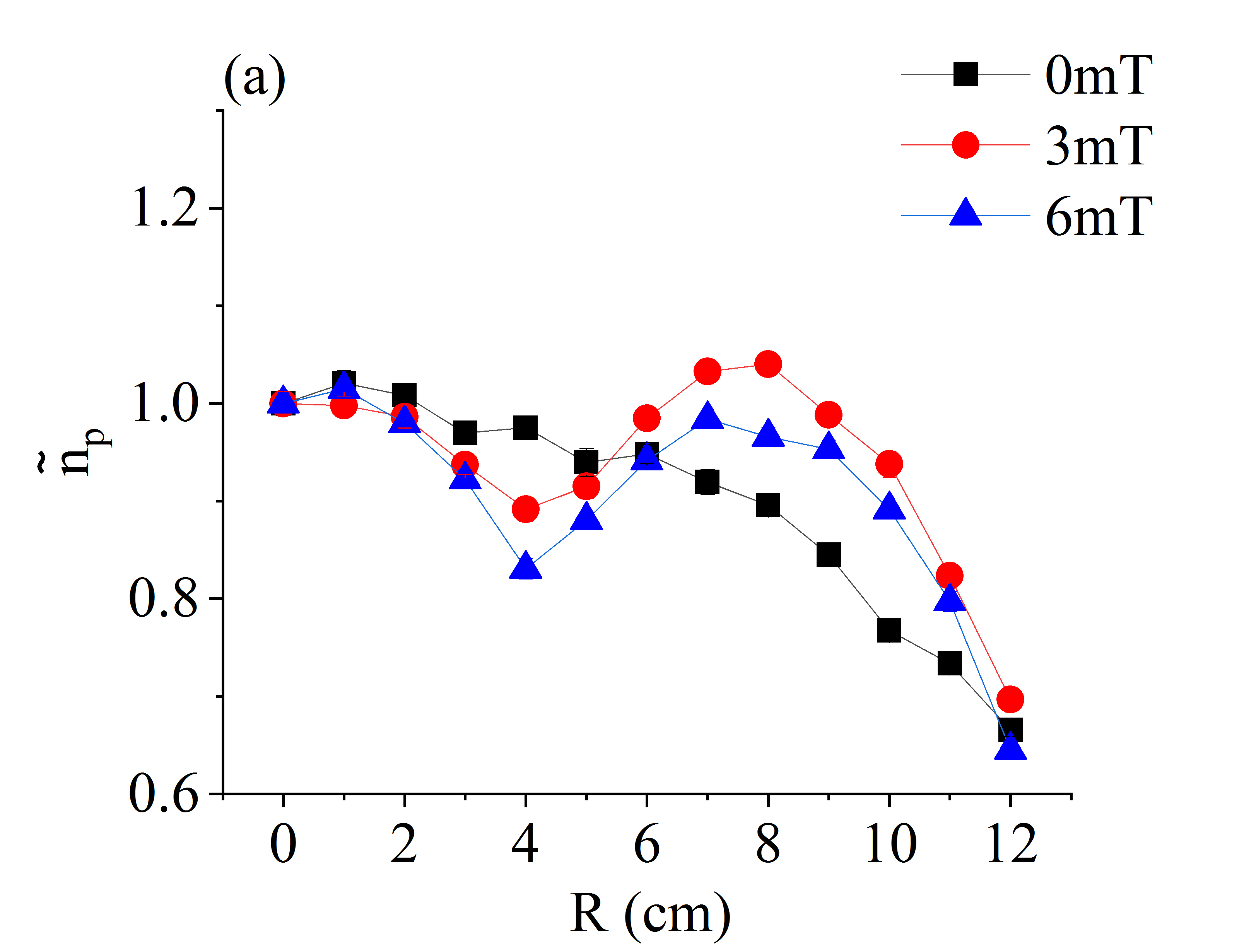}
    
  \end{minipage}
  \hfill
  \begin{minipage}[b]{0.49\textwidth}
    \includegraphics[width=\textwidth]{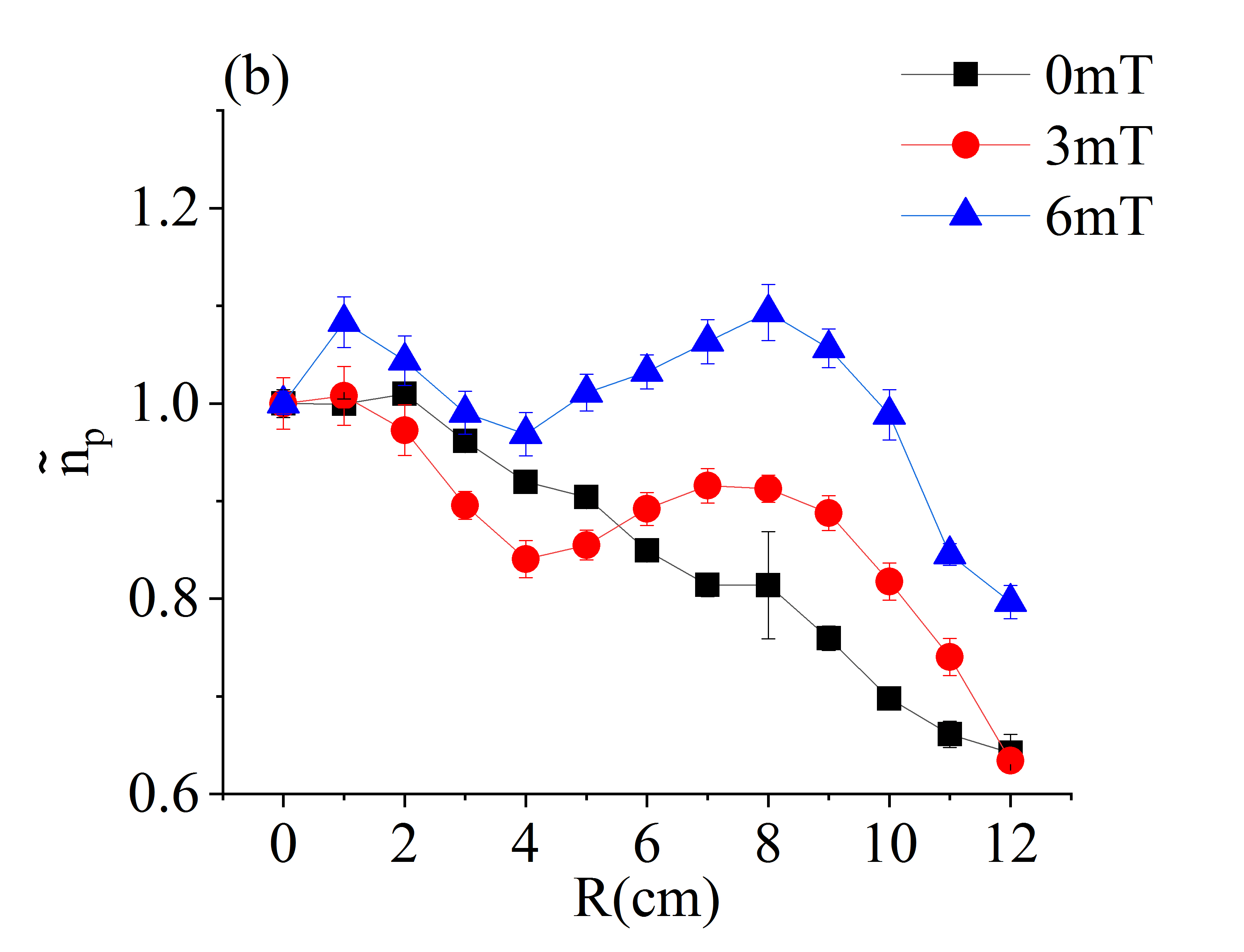}
    
  \end{minipage}
  \hfill
  \begin{minipage}[b]{0.49\textwidth}
    \includegraphics[width=\textwidth]{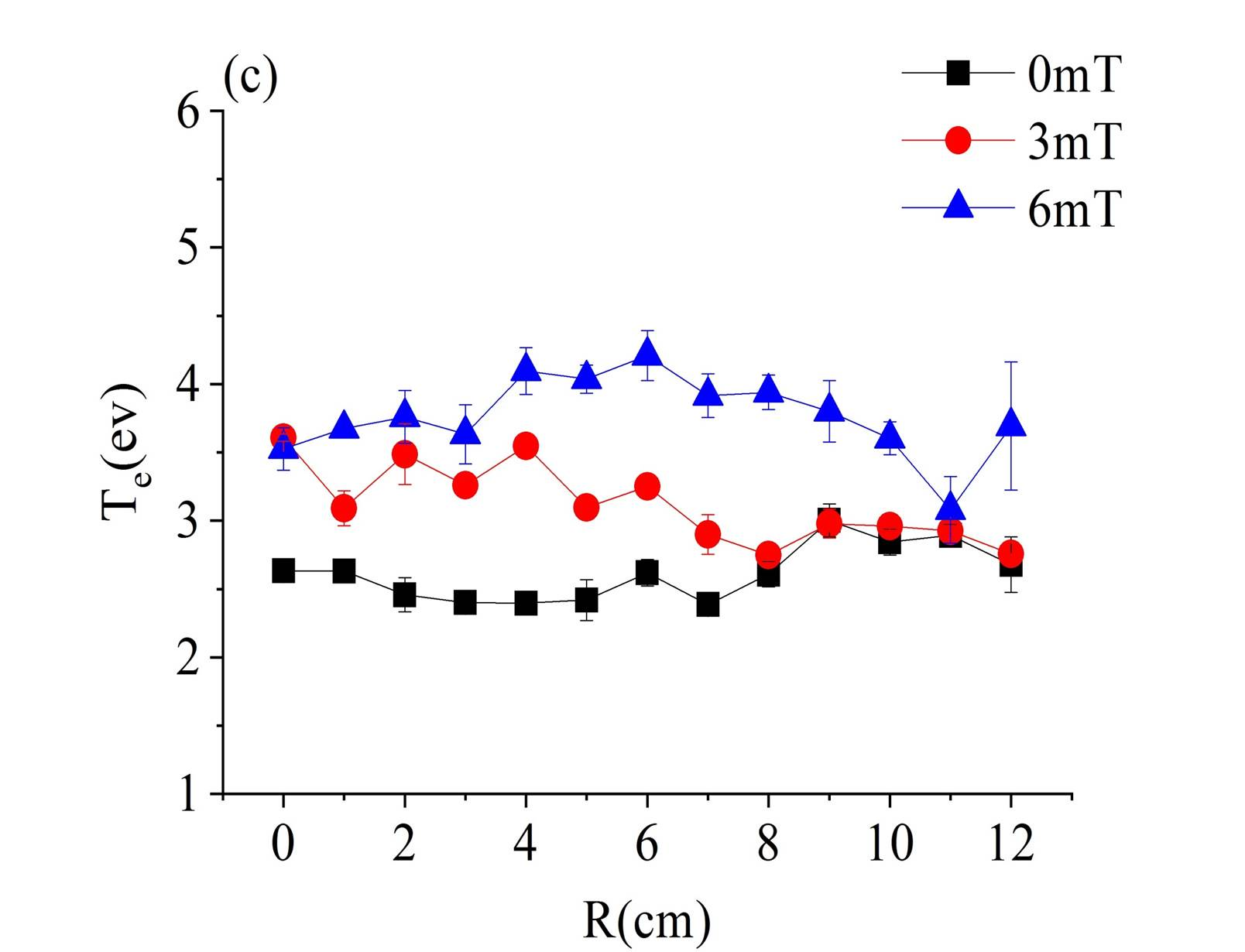}
    
  \end{minipage}
  \hfill
  \begin{minipage}[b]{0.49\textwidth}
    \includegraphics[width=\textwidth]{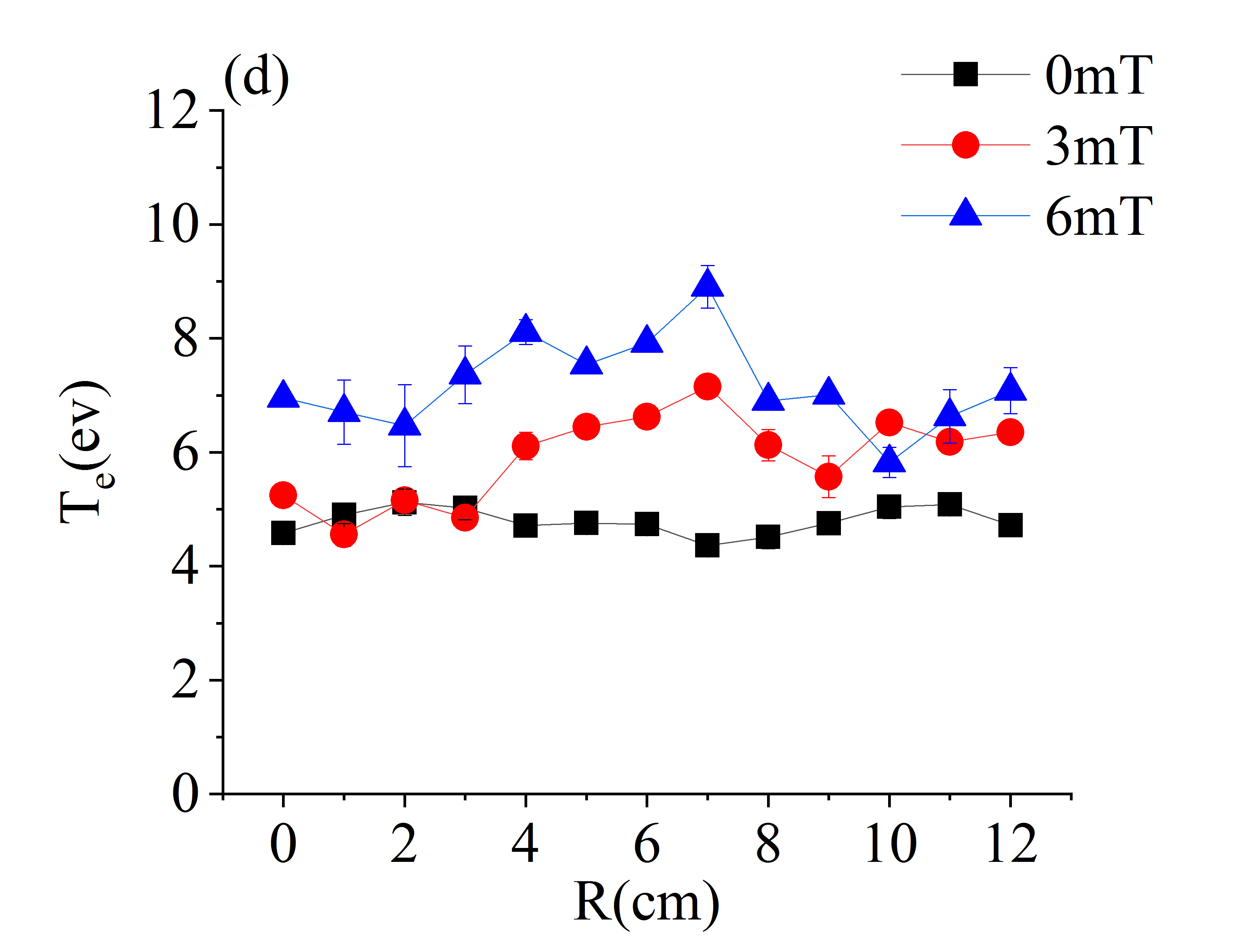}
    
  \end{minipage}
  \begin{minipage}[b]{0.49\textwidth}
    \includegraphics[width=\textwidth]{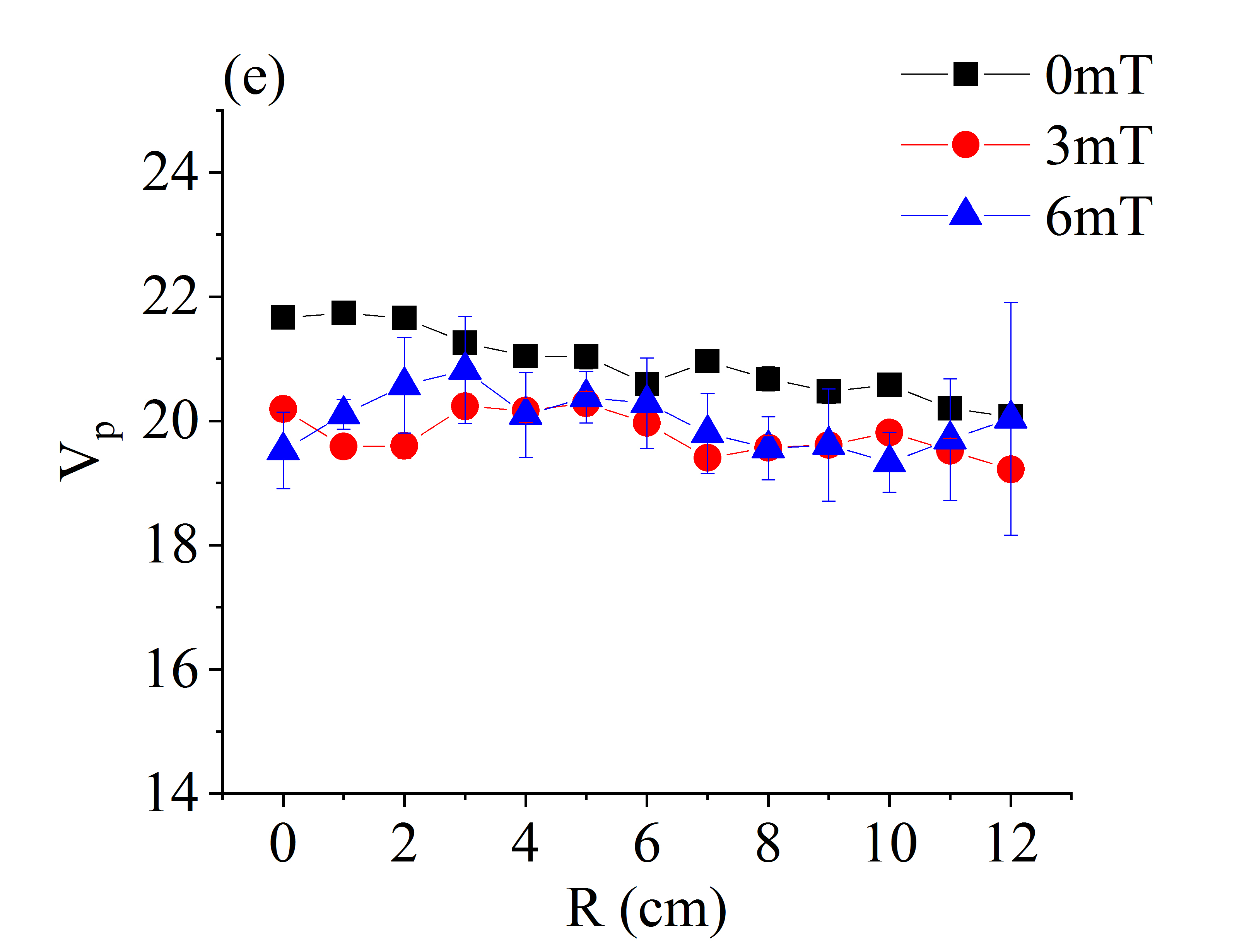}
    
  \end{minipage}
  \hfill
  \begin{minipage}[b]{0.49\textwidth}
    \includegraphics[ width=\textwidth]{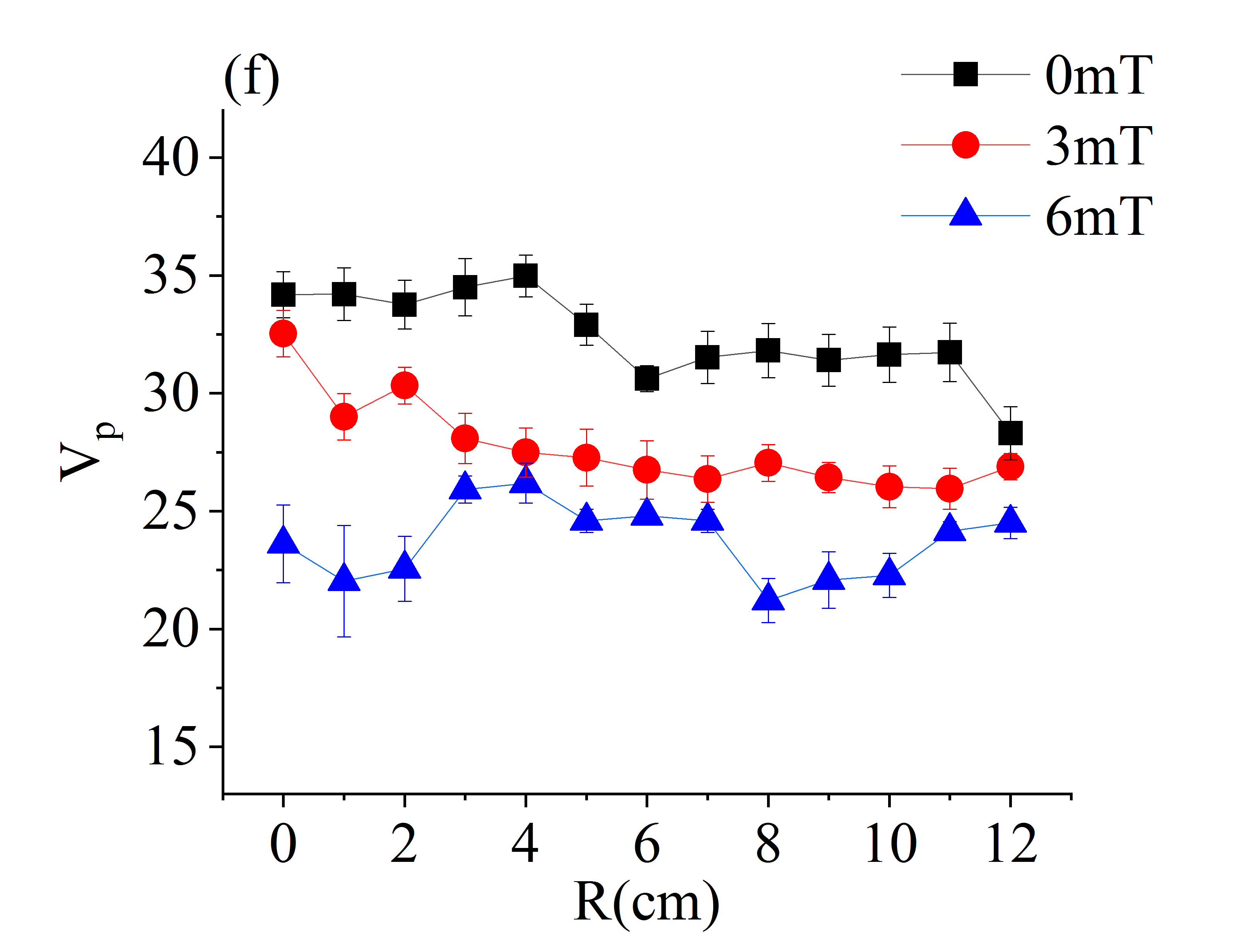}
    
  \end{minipage}
  \caption{\label{ExpMag}Radial characteristics of plasma in the presence of magnetic field (0 mT, 3mT and 6mT) at a fixed  pressure and power of 2.5 mTorr and 20 W respectively for argon (left) as well as oxygen (right) plasma . (a) Plot of radial density profile for argon with normalizing parameters 6.9249E15, 7.6726E15 and  8.3198E15 for B = 0, 3mT and 6mT  respectively, (b) plot of radial density profile for oxygen with normalizing parameters 6.14495E15, 6.1049E15 and 5.0319E15 for B = 0, 3mT and 6mT respectively, (c) plot of radial electron temperature profile for argon,(d) plot of radial electron temperature profile for oxygen, (e) plot of radial plasma potential profile for argon, (f) plot of radial plasma potential profile for oxygen.}
\end{figure}

In figure~\ref{ExpMag}, radial plots are shown for plasma density $n_p$, electron temperature $T_e$ and plasma potential $V_p$  in argon as well as electronegative oxygen discharge for a range of varying magnetic field strengths at a fixed RF power of 20 W and pressure of  2.5 mTorr. As the axial magnetic field (B = 3mT and 6mT) is applied, the motion of plasma electrons across the magnetic field lines get severely impaired ($r_{Le} \approx$ 0.8 to 4 mm $\ll$ R); however the un-magnetized positive ions ($r_{Li}\gg$ R) remain unaffected by the external magnetic field. As a result the cross field diffusion of the plasma electrons, towards the central region will be mainly governed by collisions with the background neutrals. 

It is noticed that as the magnetic field is gradually increased, an unusual double-peaked radial plasma density appears as shown in figure~\ref{ExpMag}(a) and figure~\ref{ExpMag}(b). It is also noted that $T_e$   and $n_p$   increases with the rise in magnetic field strength; however the overall magnitude of $V_p$ is seen to reduce. In the case of oxygen plasma, the electron temperature is consistently higher at all radial positions than in the case for argon.  

\begin{figure}[!hbp]
  \centering
  
    \includegraphics[width=0.6\textwidth]{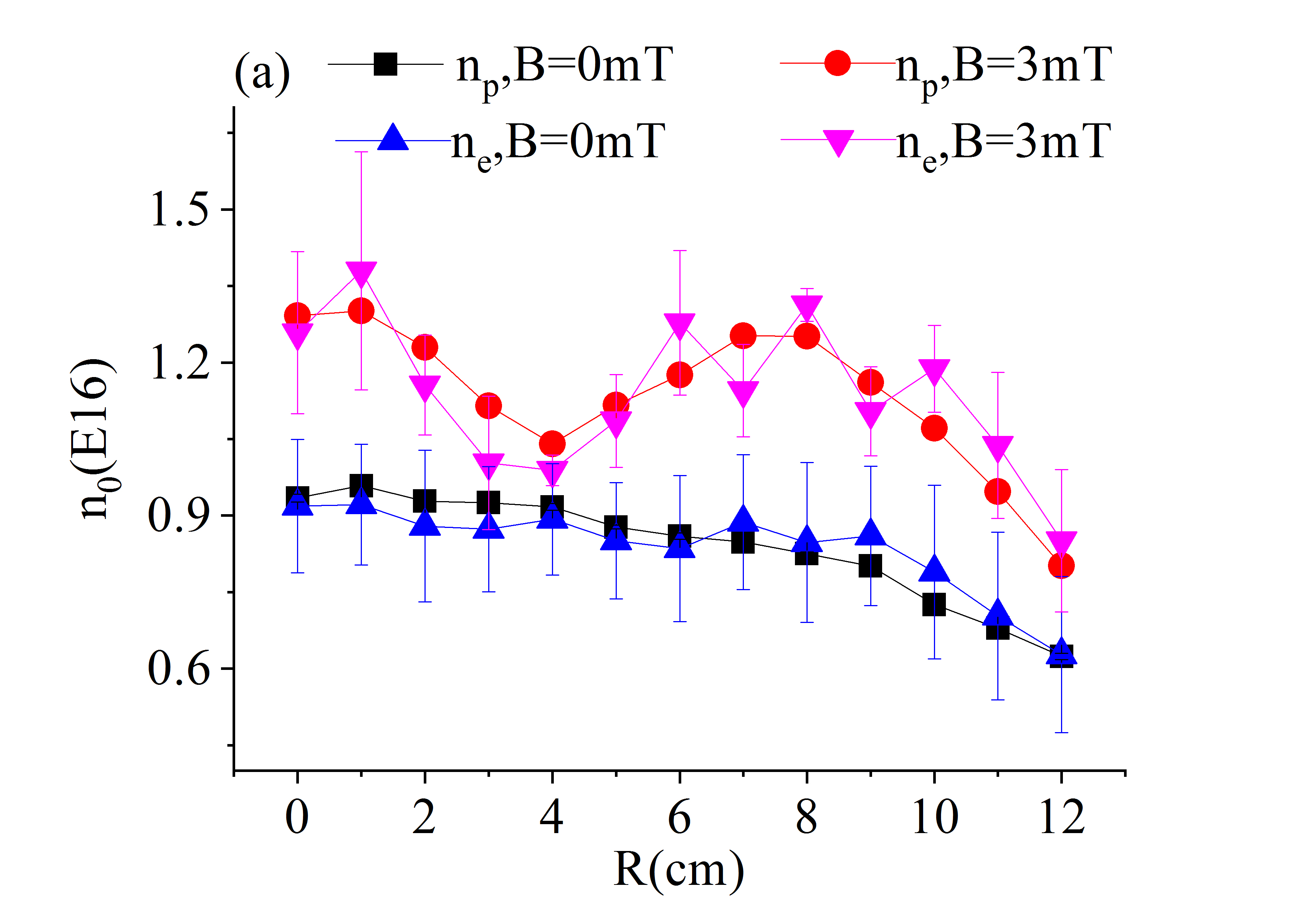}
  
    \includegraphics[width=0.6\textwidth]{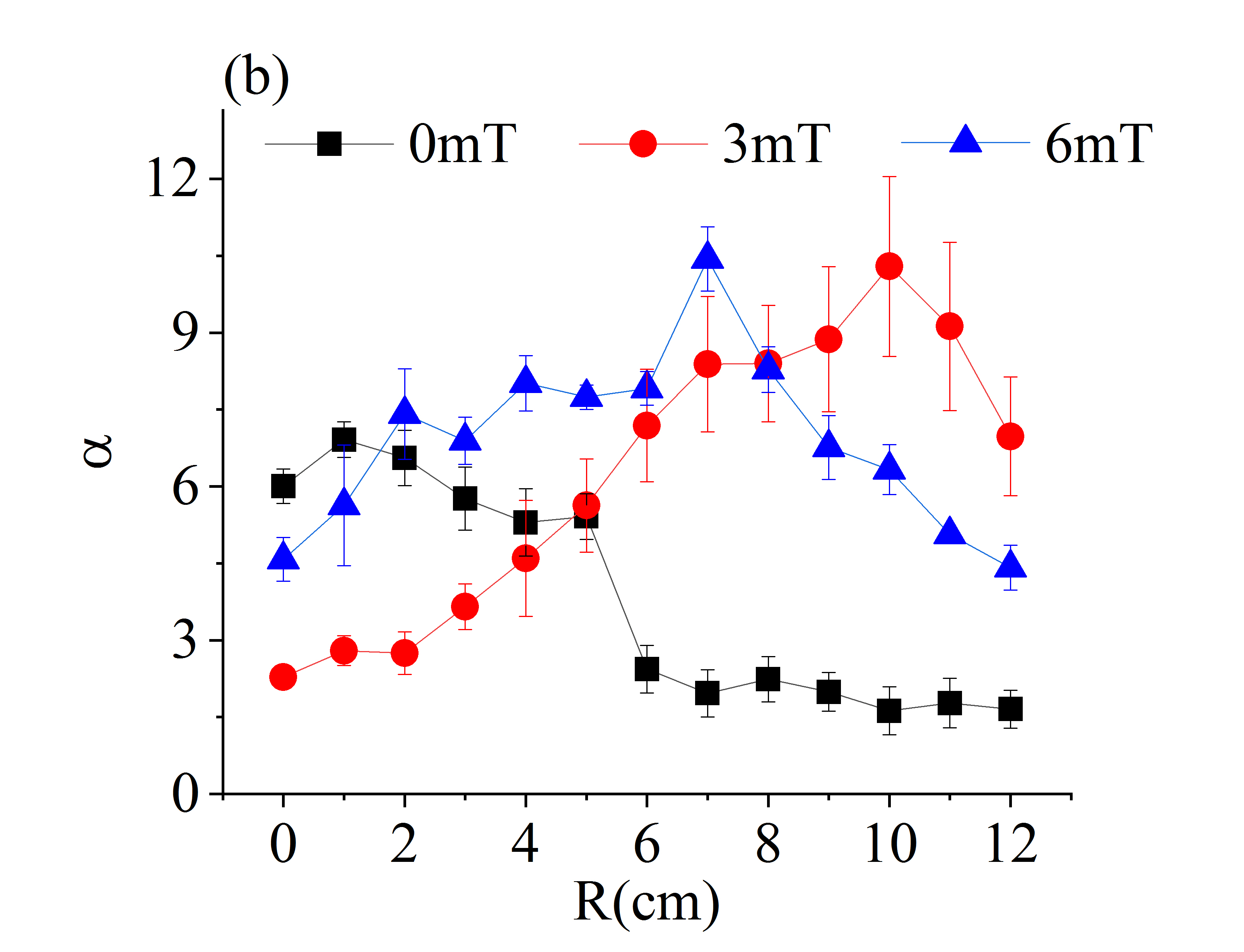}
    
    \caption{\label{nenpalpha}(a) Plot of radial number density for electron (calculated from electron saturation) and positive ions (calculated from ion saturation) for B = 0 and B=3mT , at a fixed  pressure and power of 2.5 mTorr and 20 W respectively, (b) plot of radial equilibrium distribution of $\alpha$ (ratio of negative ion density to electron density) for B = 0, 3mT and 6mT, at a fixed  pressure and power of 2.5 mTorr and 20 W respectively.}

\end{figure}

To determine the negative ion concentration inside the discharge, saturation current ratio method obtained from the Langmuir probe has been applied. The Langmuir probe has been initially calibrated in argon discharge by taking the positive ion density and electron density ratio corresponding to their respective saturation currents. As plotted in figure~\ref{nenpalpha}.a, both electron/positive ion densities match very well for the argon plasma. However in the case of oxygen plasma, plotted in figure~\ref{nenpalpha}.b, the ratio $\alpha$ (= $\frac{n_p}{n_e} -1$) varies with radial distance from the centre. The parameter $\alpha$ is defined as the electronegativity parameter (ratio of negative ion density to electron density) which gives a measure of negative ion concentration, in this case $O^-$ ions which are the most dominant species in the plasma. 

From the figure~\ref{nenpalpha}(b), it is noticed that in absence of magnetic field, there is a higher concentration of negative ions in the central region. With the application of magnetic field, the negative ions initially shift towards the edge for B = 3mT, then it becomes broader as the field is increased to B = 6mT. This observation is mainly due to volume negative $O^-$ ion; which are created by a two-step process that involves an excitation of molecular oxygen by the impact of energetic electrons and subsequently getting dissociated to form negative atomic oxygen ion by capturing an electron. 

In the absence of magnetic field, all charged species, i.e the electrons, negative ions as well as positive ions can diffuse towards the central region; however as the magnetic field increases, the inward diffusion of electrons gets reduced. This results in an off-centered peak in the plasma density. In order to maintain the quasi-neutrality, the motion of both positive as well as negative ions towards the central region is accordingly adjusted. The observation of peaked negative ion population is qualitatively higher between the capacitive plates where the excitation of oxygen species is expected to be greater due to the presence of energetic electrons. In figure~\ref{ExpMag}.f the radial plasma potential corresponding to B = 3mT attains a higher value in the central region. This potential will further repel the $O_2^+$ positive ions towards the edge where they tend to exist in equilibrium with the negative ions which already have abundance as indicated by the off-cantered peak in the electro-negativity parameter $\alpha$ in the figure~\ref{nenpalpha}(b). Summarizing the overall observations, it is found that the variation in magnetic field seems to have a strong influence on negative ion distribution and plasma density across the magnetic field; whereas only a marginal variation in electron temperature is seen in the case for oxygen. Hence in the presence of magnetic field, the peak in $\alpha$ is shifted towards the edge.

\section{\label{Model}Model}
To investigate the role of local plasma density and electron temperature on the radial characteristics of plasma potential, a radial flow model of a system consisting of electrons, positive ions and negative ions in the presence of axial magnetic field has been formulated. The model is based on fluid approximation \cite{Lieberman2005}, which takes into account the usual momentum Eq.~(\ref{FM1}), the fluid continuity Eq.~(\ref{FM2}) and combines with the charge neutrality condition Eq.~(\ref{FM3}).

\begin{eqnarray}
\fl n_j m_j (\textbf{V} _j. \nabla ) \textbf{V} _j =q_j n_j \textbf{E}+ q_j n_j (\textbf{V} _j \times \textbf{B})-\nabla(T_j n_j)- m_j \textbf{V} _j (n_j \nu_{jn}+S-L) \label{FM1}
\end{eqnarray}

\begin{equation}
$$\nabla$$.(n_j \textbf{V}_j)=S-L \label{FM2}
\end{equation}

\begin{equation}
$$\nabla$$. (\Gamma_p) = $$\nabla$$. (\Gamma_e) + $$\nabla$$. (\Gamma_v) \label{FM3}
\end{equation}

In above equations, the subscript j = e,v,p corresponds to the species electrons, volume negative ions and positive ions respectively, $T_j$  is the respective temperature of the charged species in units of eV. The collision frequency term $\nu_{jn}$ corresponds to the respective charged particle collisions with the background neutrals. S and L represent the local source and loss term respectively inside the plasma fluid while assumed to be moving with drift velocity. These collision frequencies, source and loss terms are summarized in table~\ref{RateConstants} for the all charged species in the oxygen plasma.
\begin{table}
\caption{\label{RateConstants}Collisions among the charged species in an oxygen plasma.}
\footnotesize
\begin{tabular}{@{}llll}
\br
Collisions among&Subscript name (m)&Source of&Sink of\\
\mr

positive ion-neutral                                              &$\nu_{pn}$         &ion momentum loss				&--\\
negative ion-neutrals								  &$\nu_{vn}$		&negative-ion mom. loss				&--\\
electron impact ionization                                 &$\nu_{iz}$		& electrons and ions					&neutrals\\
electron impact excitation 						 &$\nu_{ext}$		&metastable molecules				&electrons momentum\\
electron attachment								  &$\nu_{att}$		&negative ions							&electrons\\
positive-negative ions recombination   			&$\nu_{rec}$		&--								        	&positive and negative ions\\
positive ions-electron  recombination  					 &$\nu_{rec2}$		&neutrals				&electrons and positive ions\\
metastable molecule detachment		 		 &$\nu_{det}$		&de-exited molecule					&negative ions\\
electron detachment  							 &$\nu_{det2}$		&neutrals								&negative ions\\

\br
\end{tabular}\\

\end{table}
\normalsize

Using the normalized parameters

\begin{eqnarray}
\fl  N_{p,e,v}=\frac{n_{p,e,v} }{ n_0},u_{rp,rv,re}=\frac{u_{rp,rv,re} }{ c_s},\gamma_{p,v}=\frac{T_e }{ T_{p,v}},
  \mu _{pe,ve}=\frac{m_{p,v} }{ m_{e}},\mu _{pv}=\frac{m_{p} }{ m_{v}}, \xi =\frac{r }{ l_{0}},\nonumber\\
  \eta=\frac{e\phi }{ T_{e0}},\varepsilon=\frac{el_{0}E }{ T_{e0}},
 \tilde{\nu}_m=\frac{\nu_m}{c_s/l_0},c_s=\sqrt{\frac{T_e}{m_p}},\tilde{\omega}_{ce}=\frac{eB}{c_s/l_0 m_e},\label{NPs}
\end{eqnarray}
 the radial component of Eq.~(\ref{FM1}-\ref{FM2}) in cylindrical form for un-magnetized positive and negative ions, and magnetized electrons can be reduced to
 
 \begin{equation}
u_{rp}\frac{\partial u_{rp}}{\partial \xi} = \varepsilon-\frac{1}{\gamma_pN_p}\frac{\partial N_p}{\partial \xi}-\frac{u_{rp}}{N_p}(N_p \tilde{\nu}_{pn}+N_e \tilde{\nu}_{iz}-N_p \tilde{\nu}_{rec})\label{REP1}
\end{equation}

 \begin{equation}
\frac{1}{\xi}\frac{\partial (\xi N_p u_{rp})}{\partial \xi} = N_e \tilde{\nu}_{iz}-N_p \tilde{\nu}_{rec}\label{REP2}
\end{equation}

 \begin{equation}
\fl u_{rv}\frac{\partial u_{rv}}{\partial \xi} = -\mu _{pv} \varepsilon-\frac{\mu _{pv}}{\gamma_v N_v}\frac{\partial N_v}{\partial \xi}-\frac{u_{rv}}{N_v}(N_v \tilde{\nu}_{vn}+N_e \tilde{\nu}_{att}-N_v \tilde{\nu}_{rec}-N_v \tilde{\nu}_{det}-N_v \tilde{\nu}_{det2})\label{REN1}
\end{equation}

 \begin{equation}
\frac{1}{\xi}\frac{\partial (\xi N_v u_{rv})}{\partial \xi} = N_e \tilde{\nu}_{att}-N_v \tilde{\nu}_{rec}-N_v \tilde{\nu}_{det}-N_v \tilde{\nu}_{det2}\label{REN2}
\end{equation}

 \begin{equation}
\fl 0 = -\mu _{pe} \varepsilon-\frac{\tilde{\omega}^{2}_{ce} u_{er}}{(\tilde{\nu}_{ext}+\tilde{\nu}_{iz}-\tilde{\nu}_{att})}-\frac{\mu _{pe}}{ N_e T_e}\frac{\partial N_e T_e}{\partial \xi}-u_{re}(\tilde{\nu}_{ext}+\tilde{\nu}_{iz}-\tilde{\nu}_{att})\label{REE1}
\end{equation}

 \begin{equation}
\frac{1}{\xi}\frac{\partial  (\xi N_e u_{re})}{\partial \xi} = N_e \tilde{\nu}_{ext}+N_e\tilde{\nu}_{iz}-N_e\tilde{\nu}_{att}\label{REE2}
\end{equation}
where the parameters, $n_0$ and $T_{e0}$ are respectively the positive ion density and the electron temperature at the centre of the discharge and $c_s$ is the positive ion bohm speed. Using Eq.~\ref{REP1} and Eq.~\ref{REP2}, we can express the radial flux of positive ions as;

 \begin{equation}
\Gamma_{pr} = \frac{1}{\tilde{\nu}_{pn}} [ (u^{2}_{rp} -\frac{1}{\gamma_p})\frac{\partial N_p}{\partial \xi}+N_p  \varepsilon+\frac{N_pu^{2}_{rp}}{\partial \xi} ] \label{FP}
\end{equation}

Similarly using Eq.~\ref{REN1} and Eq.~\ref{REN2}, we can write the flux of negative ions;

 \begin{equation}
\Gamma_{vr} = \frac{1}{\tilde{\nu}_{vn}}[(u^{2}_{rv} -\frac{\mu _{pv}}{\gamma_v})\frac{\partial N_v}{\partial \xi}+\mu _{pv}N_v  \varepsilon+\frac{N_v u^{2}_{rv}}{\partial \xi}]\label{FN}
\end{equation}

and using Eq.~\ref{REE1} and Eq.~\ref{REE2}, the electron flux is given by;

 \begin{equation}
\Gamma_{er} = \frac{1}{\tilde{\nu}_{T}}[\frac{\mu _{pe}}{T_e}\frac{\partial (N_e T_e)}{\partial \xi}+\mu _{pe}N_e  \varepsilon]\label{FE}
\end{equation}

where $\tilde{\nu}_{T}=(\tilde{\nu}_{ext}+\tilde{\nu}_{iz}-\tilde{\nu}_{att})\frac{\tilde{\omega}^{2}_{ce} }{(\tilde{\nu}_{ext}+\tilde{\nu}_{iz}-\tilde{\nu}_{att})^{2}}$. In the present setup, the capacitive discharge is created in a push-pull configuration via an isolation transformer. This geometry ensures that a zero mean current at both the electrodes which independently satisfy  the flux balance in both axial as well as radial direction. Hence the grounded end-plates has no role in producing a short-circuiting effect. Therefore, the quasi-neutrality condition at each radial location requires that the net flux of positive and negative charges should balance at all radial position. Therefore,

 \begin{equation}
\Gamma_{pr} = \Gamma_{vr}+\Gamma_{er}\label{FB}
\end{equation}

Also, positive ions can be assumed to have an isotropic temperature inside bulk plasma; hence it is convenient to approximate the radial velocity of positive and negative ions to $u^{2}_{rp}=\frac{2}{\gamma_p}$  and $u^{2}_{rv}=\frac{2\mu_{pv}}{\gamma_v}$  respectively. Therefore, using Eq.~(\ref{FP}) to Eq.~(\ref{FE}) in Eq.~(\ref{FB}) , and defining $\alpha=N_v/N_e$   ,we can write

\begin{eqnarray}
\fl \varepsilon (1+\frac{\tilde{\nu}_{pn}}{\tilde{\nu}_{vn}} \frac{\alpha \mu_{pv}}{1+\alpha}+\frac{\tilde{\nu}_{pn}}{\tilde{\nu}_T} \frac{ \mu_{pe}}{1+\alpha})=\frac{1}{N_p} \frac{\tilde{\nu}_{pn}}{\tilde{\nu}_{vn}} \frac{\mu_{pv}}{\gamma_v} \frac{\partial(\frac{\alpha}{1+\alpha}N_p)}{\partial	\xi}+\frac{\alpha}{1+\alpha}\frac{\tilde{\nu}_{pn}}{\tilde{\nu}_{vn}} \frac{\mu_{pv}}{\xi \gamma_v}-\nonumber\\ \frac{1}{\gamma_p N_p}\frac{\partial N_p}{\partial \xi}-\frac{1}{\gamma_p \varepsilon}-\frac{\tilde{\nu}_{pn}}{\tilde{\nu}_{T}} \frac{\mu_{pe}}{N_p T_e} \frac{\partial(\frac{N_p}{1+\alpha}T_e)}{\partial	\xi}\label{TM1}
\end{eqnarray}
or,

\begin{eqnarray}
\fl \varepsilon (1+\frac{\tilde{\nu}_{pn}}{\tilde{\nu}_{vn}} \frac{\alpha \mu_{pv}}{1+\alpha}+\frac{\tilde{\nu}_{pn}}{\tilde{\nu}_T} \frac{ \mu_{pe}}{1+\alpha})=\frac{1}{N_p} \frac{\tilde{\nu}_{pn}}{\tilde{\nu}_{vn}} \frac{\mu_{pv}}{\gamma_v}\frac{\alpha}{1+\alpha} \frac{\partial(N_p)}{\partial	\xi}+\nonumber\\\frac{\tilde{\nu}_{pn}}{\tilde{\nu}_{vn}} \frac{\mu_{pv}}{\gamma_v} \frac{\partial(\frac{\alpha}{1+\alpha})}{\partial	\xi}+\frac{\alpha}{1+\alpha}\frac{\tilde{\nu}_{pn}}{\tilde{\nu}_{vn}} \frac{\mu_{pv}}{\xi \gamma_v}-\frac{1}{\gamma_p N_p}\frac{\partial N_p}{\partial \xi} -\nonumber\\ \frac{1}{\gamma_p \varepsilon}-\frac{\tilde{\nu}_{pn}}{\tilde{\nu}_{T}} \frac{\mu_{pe}}{N_p } \frac{1}{1+\alpha}\frac{\partial N_p}{\partial	\xi}-\frac{\tilde{\nu}_{pn}}{\tilde{\nu}_{T}} \frac{\mu_{pe}}{ T_e} \frac{1}{1+\alpha}\frac{\partial T_e}{\partial	\xi}-\frac{\tilde{\nu}_{pn}}{\tilde{\nu}_{T}} \mu_{pe} \frac{\partial(\frac{1}{1+\alpha})}{\partial \xi}\label{TM2}
\end{eqnarray}

where $N_v= N_p\alpha/(1+\alpha)$ and$ N_e=N_p/(1+\alpha)$. Since for non-thermal oxygen plasma where $T_e \gg T_{p,v}$, $1/\gamma_v =1/\gamma_p \ll 1$ and $\mu_{pv} (\approx 0.5)\ll \mu_{pe}(\approx 10^4)$, and by assuming $\tilde{\nu}_{pn}=\tilde{\nu}_{vn}$, $\tilde{\nu}_{T}\gg \tilde{\nu}_{pn}$, which is indeed true for low pressures and partially magnetized plasma ($\tilde{\nu}_{T} \approx 10^{10-11}\gg \tilde{\nu}_{pn} \approx 10^{8-9}$), the above Eq.~{\ref{TM2}} reduces to

\begin{equation}
\fl \varepsilon (\frac{\tilde{\nu}_{pn}}{\tilde{\nu}_T} \frac{ \mu_{pe}}{1+\alpha})=\frac{\tilde{\nu}_{pn}}{\tilde{\nu}_{T}} \frac{\mu_{pe}}{N_p } \frac{1}{1+\alpha}\frac{\partial N_p}{\partial	\xi}-\frac{\tilde{\nu}_{pn}}{\tilde{\nu}_{T}} \frac{\mu_{pe}}{ T_e} \frac{1}{1+\alpha}\frac{\partial T_e}{\partial	\xi}-\frac{\tilde{\nu}_{pn}}{\tilde{\nu}_{T}} \mu_{pe} \frac{\partial(\frac{1}{1+\alpha})}{\partial	\xi}\label{TM3}
\end{equation}

Therefore the normalized electric field ($\varepsilon$) can be expressed as,

\begin{equation}
-\varepsilon= \frac{\partial (\phi/T_e)}{\partial \xi} =\frac{\partial (lnN_p)}{\partial \xi}+\frac{\partial (lnT_e)}{\partial \xi}+\frac{\partial (ln(\frac{1}{1+\alpha}))}{\partial \xi}\label{TM4}
\end{equation}

Integration of  Eq.~{\ref{TM4}} provides a radial potential in terms of normalized positive ion density $ N_p$, electron temperature $T_e$  and electronegativity parameter $\alpha$.

\begin{equation}
\phi =T_e ln(\frac{N_p T_e}{1+\alpha})\label{TM5}
\end{equation}

For the electropositive argon case, $\alpha=0 $ i.e., therefore the above equation reduces to

\begin{eqnarray}
\phi =T_e ln(N_p T_e)\label{TM6}
\end{eqnarray}

It is important to mention that the above analytical form is possible by making the above approximations,  $\tilde{\nu}_{pn}=\tilde{\nu}_{vn}$, $\tilde{\nu}_{T}\gg \tilde{\nu}_{pn}$; which is valid for a low pressure plasma. However for the high pressure collision systems, these approximations may not be valid, hence one has to seek to solve the Eq.~{\ref{TM2}} numerically to obtain the radial potential profile of the system.

\section{\label{ResDis}Results and Discussion:}
\begin{figure}[!hbp]
  \centering
  \begin{minipage}[b]{0.49\textwidth}
    \includegraphics[width=\textwidth]{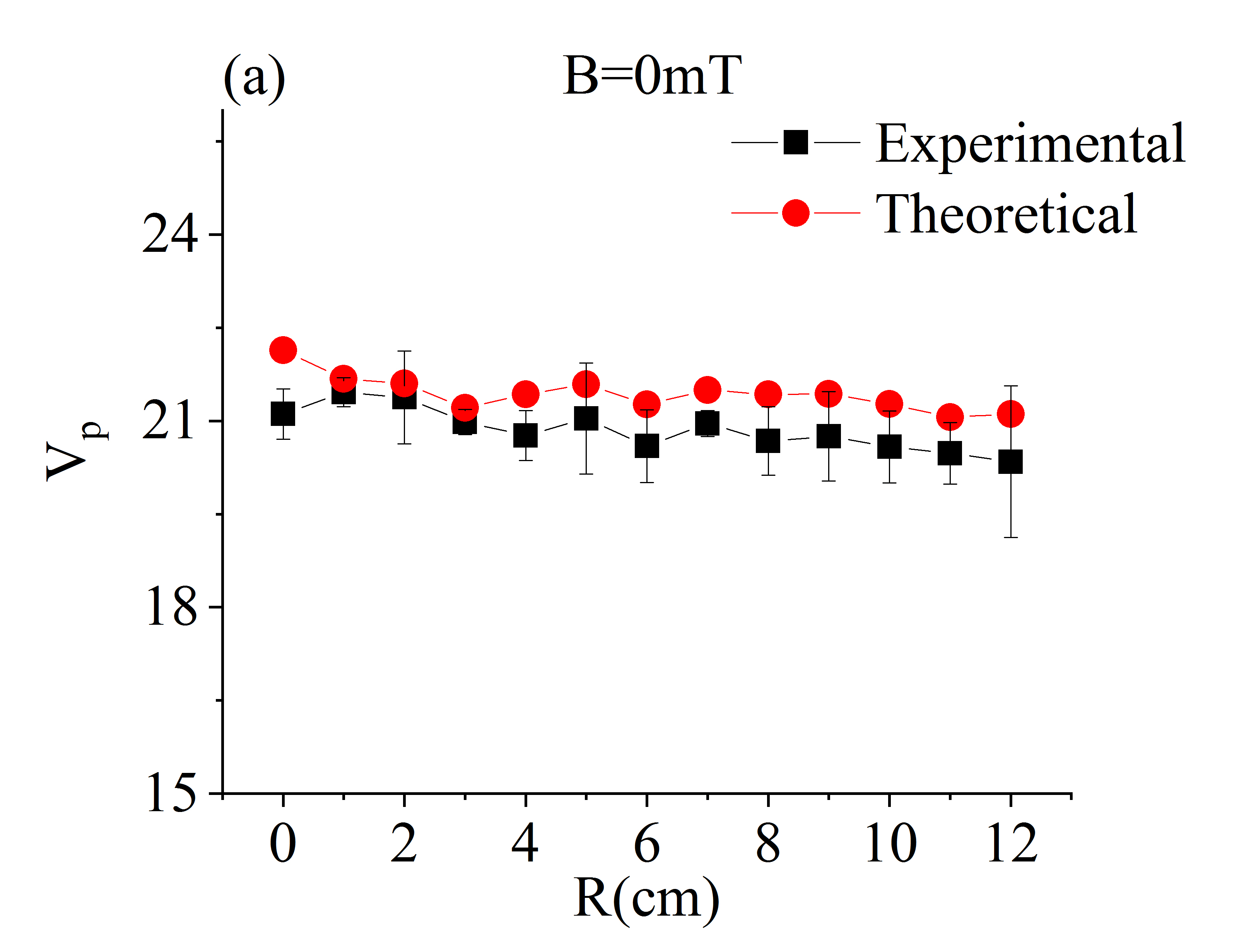}
   
  \end{minipage}
  \hfill
  \begin{minipage}[b]{0.49\textwidth}
    \includegraphics[width=\textwidth]{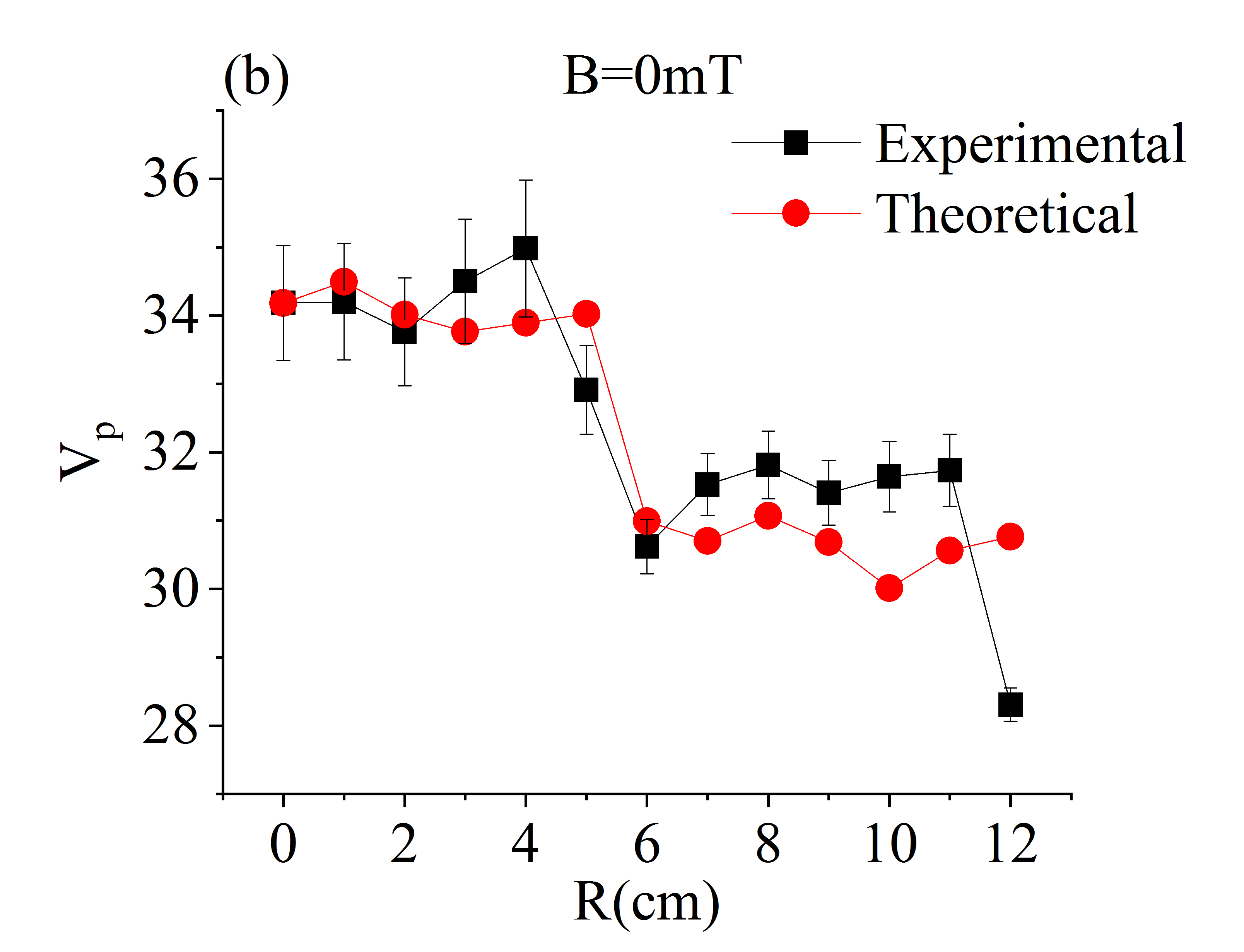}
    
  \end{minipage}
  \begin{minipage}[b]{0.49\textwidth}
    \includegraphics[width=\textwidth]{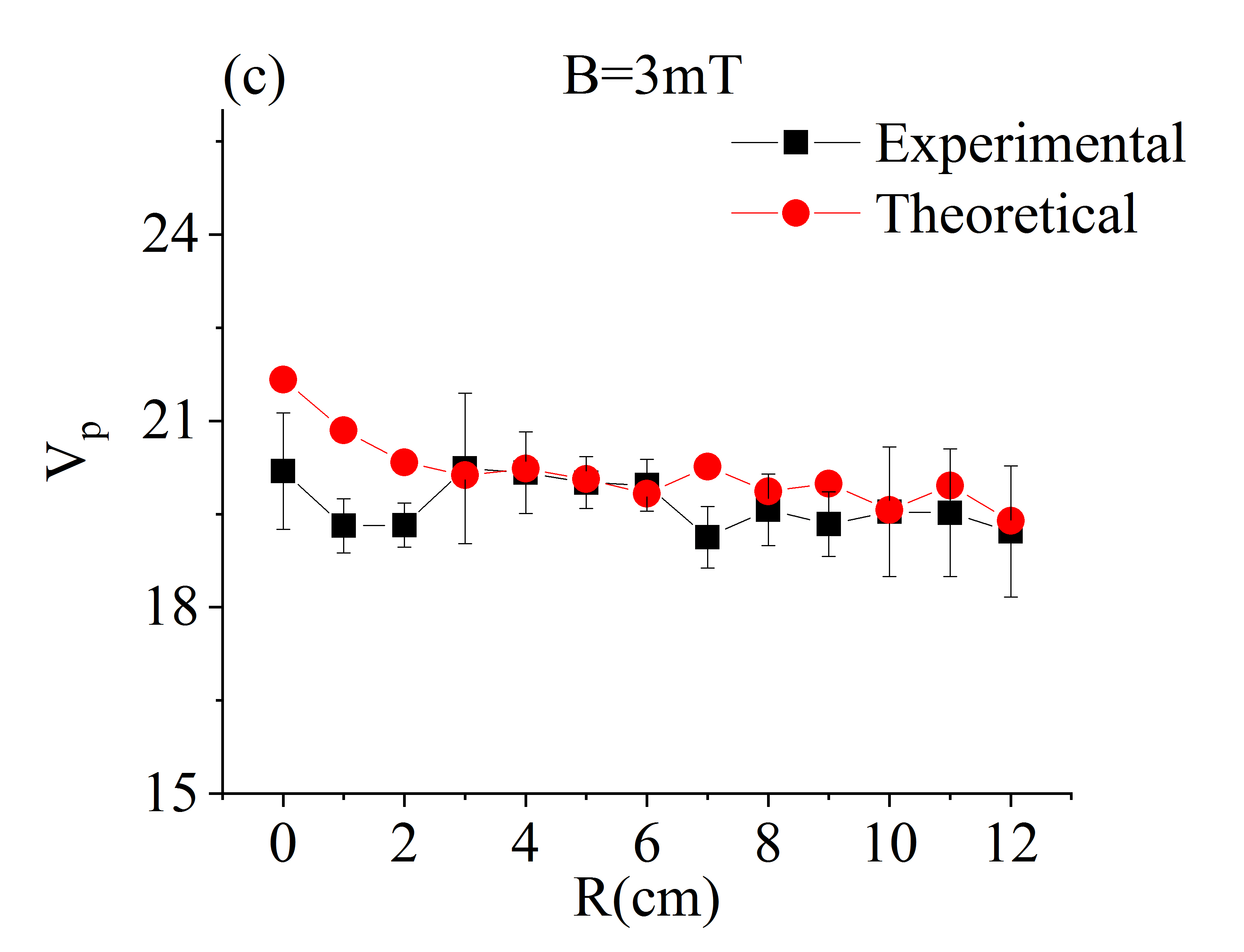}
   
  \end{minipage}
  \hfill
  \begin{minipage}[b]{0.49\textwidth}
    \includegraphics[width=\textwidth]{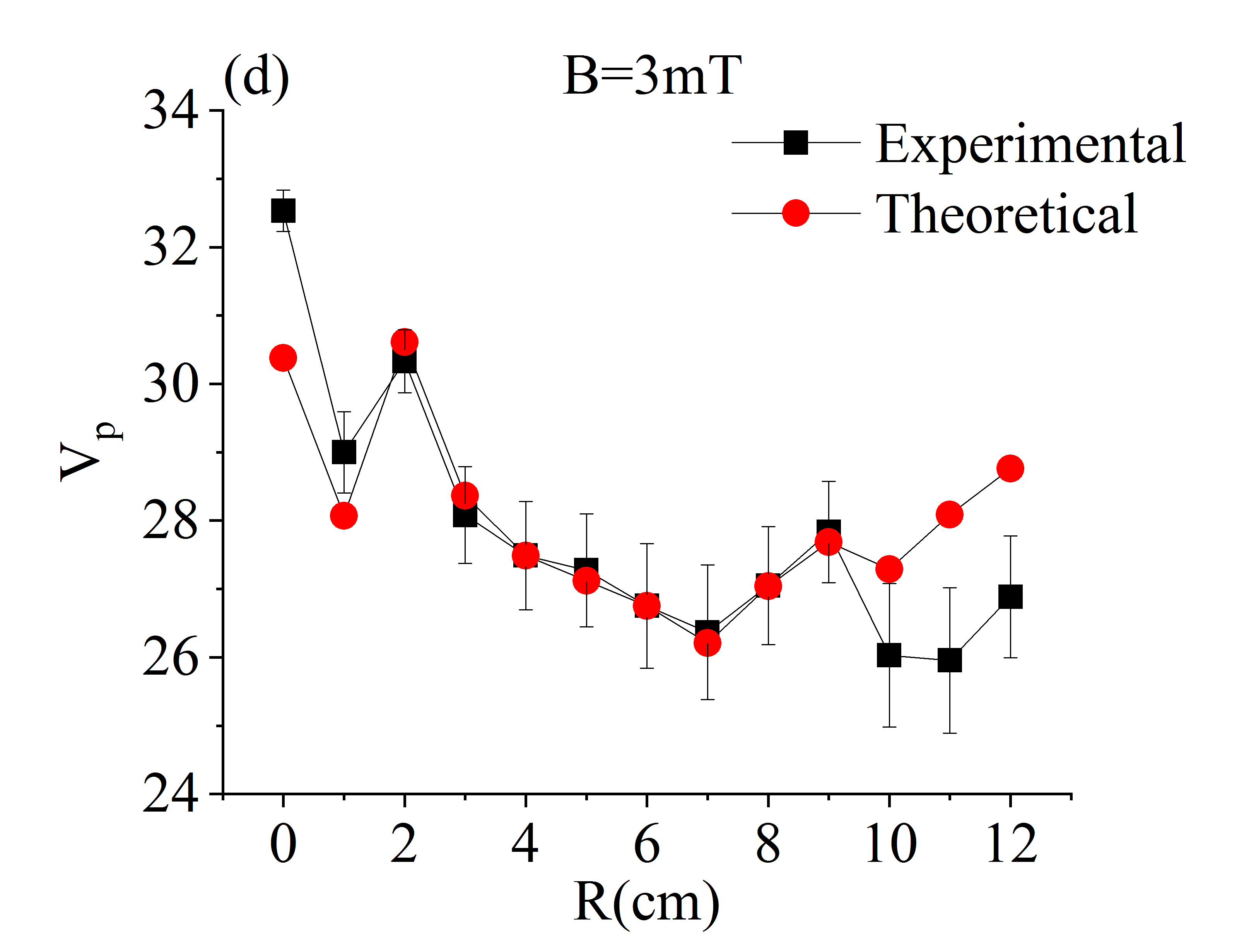}
    
  \end{minipage}
  \begin{minipage}[b]{0.49\textwidth}
    \includegraphics[width=\textwidth]{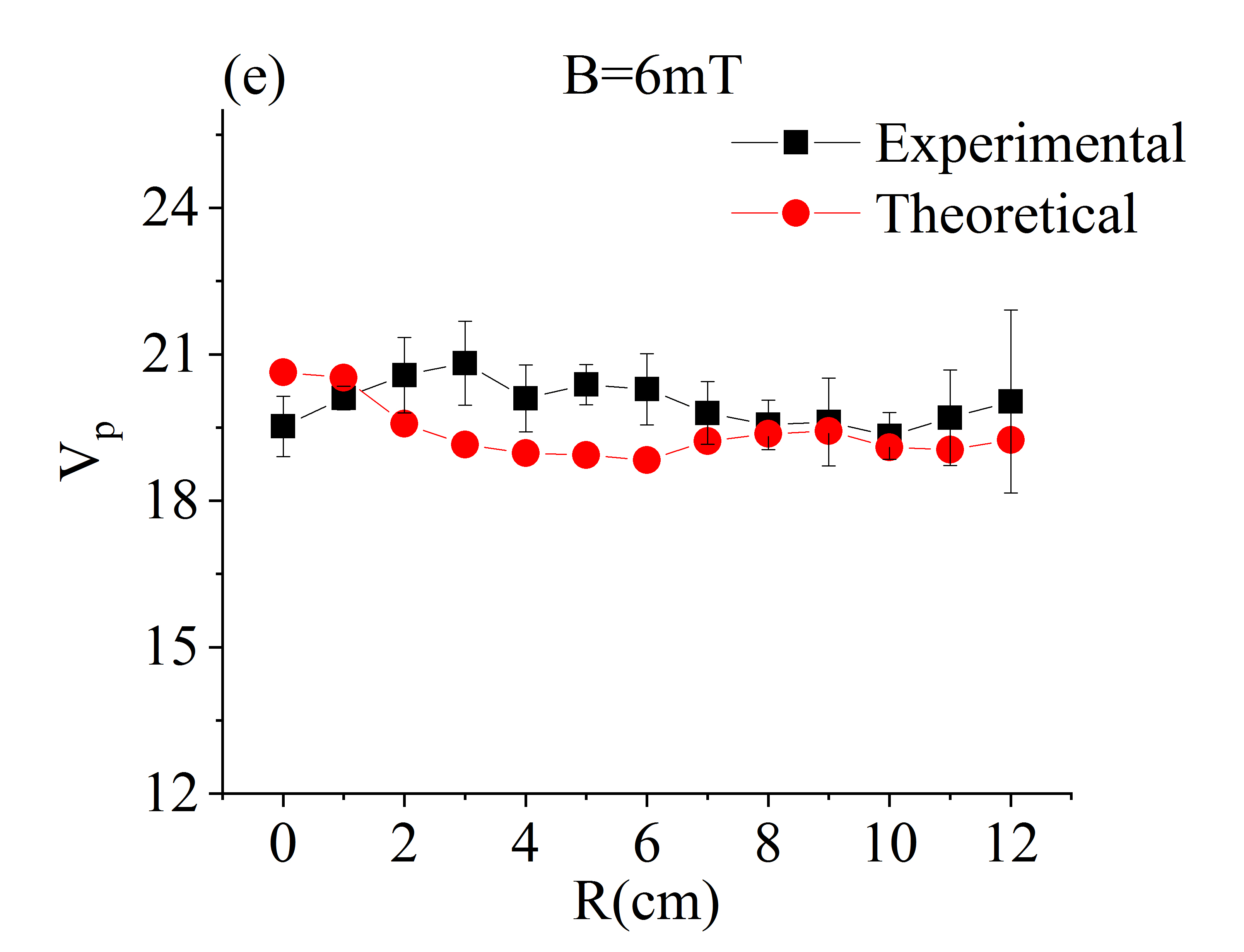}
    
  \end{minipage}
  \hfill
  \begin{minipage}[b]{0.49\textwidth}
    \includegraphics[width=\textwidth]{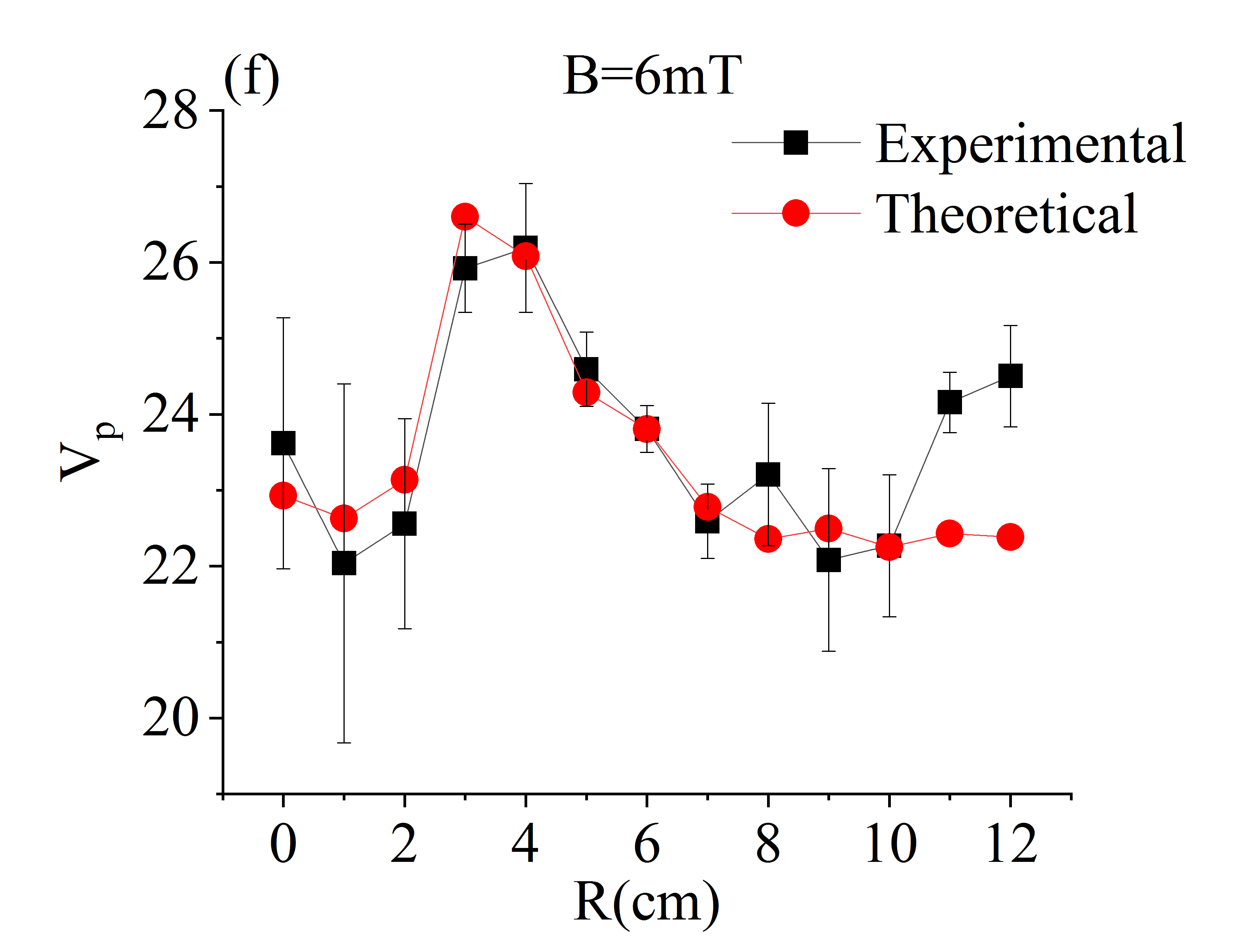}
    
  \end{minipage}
   \caption{\label{ExpThr}Plot of theoretical and experimental radial plasma potential profile  for the argon (left) and oxygen (right) plasma having B = 0mT, 3mT and 6mT, at a fixed  pressure and power of 2.5 mTorr and 20 W respectively. Plot of radial plasma potential profile for (a) argon at B=0 , (b) oxygen at B = 0, (c)  argon at B  = 3mT, (d) oxygen at B = 3mT, (e) argon at B = 6mT, (f) oxygen at B = 6mT.}
\end{figure}

Using the above equations, the radial plasma potential in the mid-plane between capacitive plates has been obtained for argon and as well as the oxygen case under varying magnetic field strengths as plotted in figures~\ref{ExpThr}. A very good agreement is seen with the experimental trend and from this model. It is interesting to note that the magnetic field parameter does not appear directly in the formula of plasma potential ($V_p$), but the radial plasma density and electron temperature profile is already influenced by the external magnetic field.  

This model basically tells that in the case of inhomogeneous plasma, which has a radial variation in electron temperature as well as in plasma density; both collectively determines the plasma potential. Furthermore, relation between the radial plasma potential, density and temperature is analogous with the Boltzmann relation observed for un-magnetized case which is also the case of an infinitely long plasma column along an axial magnetic field line. However the model does not tells about the main driving force among the three candidates, which is responsible for the transport of charged particles across the magnetic field lines. Nevertheless, the model can predict the radial potential distribution due to the presence of negative ions. 

In figure~\ref{ExpMag}(a), the region between the plates (off-centered region) must have a significant population of hotter electrons due to rapidly oscillating electrons from the opposing sheaths. This is also confirmed from the radial plots of $T_e$   which shows an increasing trend; however, the low energy electrons efficiently makes a random walk across the magnetic field lines to reach the central region as compared to the hotter population. The cross-field moving electrons are followed by positive ions to provide them charge shielding. As a result the central region for both argon and oxygen plasma continues to observe a peak in density, irrespective of increase in the magnetic field strength. Therefore with the increase in magnetic field a double hump density structure is observed. 

In the case of oxygen, $O^-$  ions are formed from $O_2$ via dissociative attachment of electrons to the oxygen atoms. It is more likely that the dissociation rate is expected to be higher in the region between the plates due to the presence of energetic population of electrons. Therefore the probability of negative ion formation in the outer region is higher. In figure~\ref{ExpMag}(b) the central plasma density consistently remains  higher, irrespective of increase in magnetic field strength as low energy electrons are accompanied along with $O_2^+$ ions. As a result, the central plasma region tends towards having an electropositive core than the outer region.  It seems that in this particular case, two distinct plasma regions are created in presence of magnetic field; one is the central region with electron and positive ion plasma and the peripheral region surrounded by the more dominant $O^-$  ions. This is consistent with the plot of $\alpha$ with radial distance in figure~\ref{nenpalpha}(b). 

In certain apparatus, radial potential distribution across the magnetic field can be affected in the presence of a conducting end-plate \cite{Das2019a,Schulze2008,Schwirzke1964}. Whereas in the present case, the electrodes geometry does not prefers a short-circuit effect but rather shows independent flux balance in the axial (as confirmed by the zero mean current over a RF cycle) as well in the radial direction (as also confirmed by this radial flow fluid model). As a matter of fact the model cannot explain the experimental data reported in \cite{Das2019a}, which was the case when the plasma body was in contact with the grounded chamber wall, thus providing a short-circuiting path for the electrons through the metallic body of the chamber. Thus the bottom-line is that the radial potential, density and temperature in an inhomogeneous plasma system created due to application of magnetic field, pressure or distributed sources will follow a Boltzmann like distribution, provided that the conducting plasma boundaries do not influence the potential profile inside the plasma. 

\section{\label{Sum}Summary:}
This paper describes an analytical model to calculate the radial potential distribution of a magnetized CCRF discharge which is associated with strong variation in plasma density, electronegativity parameter and electron temperature. The density profile did not followed the usual plasma potential as expected for the case of a Boltzmann distribution of electrons, neither the phenomenological model with incorporating the effect due to conducting end plate could match the experiments. Furthermore the potential distribution of partially magnetized plasma in presence of negative ions has not been explored in the past. 

This paper addressed the above problem by showing that under the approximation when the axial flow of plasma to the wall can be ignored or if it independently balance the flux of positive and negative charged species towards the axial boundaries, then this radial flux balance model leads to a simpler expression of plasma potential as a function of electron density, electron temperature and the bulk electronegativity parameter. This model has been tested in the present setup and is found to be in good qualitative agreement for collision-less low pressure system. The model can be helpful to predict the spatial distribution of plasma parameter for example in the filter field region inside ion sources.   
 
 \ack
 The authors thank Dr Mainak Bandyopadhyay for his valuable comments/suggestions regarding the manuscript. This work is supported by the Department of Atomic Energy, Government of India.

\section*{References}

\bibliographystyle{unsrt}
\bibliography{Source_Paper}
\pagebreak

\pagebreak

\end{document}